\documentstyle[psfig]{aipproc}

\begin{document}
\title{The Linear $U3\times U3$ Sigma Model, the $\sigma(\approx 500)$
and  the Spontaneous Breaking of Symmetries\footnote{Talk given at
Hadron97 or
International Conference on Hadron Spectroscopy, Brookhaven, USA Aug
24-30 1997. And at the EuroDaphne meeting in Barcelona Nov. 6-9 1997.} }

\author{Nils A. T\"ornqvist}

\address{Physics Department, 
POB 9, FIN--00014, University of Helsinki, Finland}

\maketitle

\begin{abstract}
My recent fit of the light scalar mesons is discussed in the light of the
$U3\times U3$ linear sigma model. The argument for why there exists a light
and broad $\sigma(\approx 500)$  meson is explained in more detail, and the
mechanism of spontaneous symmetry breaking is discussed.
\end{abstract}

\def \gam {\frac{ N_f N_cg^2_{\pi q\bar q}}{8\pi} }
\def \gamm {N_f N_cg^2_{\pi q\bar q}/(8\pi) }
\def \be {\begin{equation}}
\def \ba {\begin{eqnarray}}
\def \ee {\end{equation}}
\def \ea {\end{eqnarray}}
\def \gap {{\rm gap}}
\def \gapp {{\rm \overline{gap}}}
\def \gappp {{\rm \overline{\overline{gap}}}}
\def \im {{\rm Im}}
\def \re {{\rm Re}}
\def \Tr {{\rm Tr}}
\def \P {$0^{-+}$}
\def \S {$0^{++}$}
\def \uu {$u\bar u+d\bar d$}
\def \ss {$s\bar s$}

%\section{Introduction}

The first thing I want to point out is that my recent model fit to the
light scalar $q \bar q$ nonet \cite{NAT}, in fact, can be interpreted as
a model unitarization based  upon the  $U(3) \times U(3)$ 
chiral symmetric renormalizable Lagrangian
\be  {\cal L}=
\frac 1 2 \Tr [\partial_\mu\Sigma \partial_\mu\Sigma^\dagger]
+\frac 1 2 m^2\Tr [\Sigma \Sigma^\dagger] -\lambda \Tr[\Sigma\Sigma^\dagger
\Sigma\Sigma^\dagger]  -\lambda' (\Tr[\Sigma\Sigma^\dagger])^2 
+{\cal L}^{SB} \ .
\label{lag}
\ee
Here $\Sigma= \sum_{i=0}^8(s_i+ip_i)\lambda_i/\sqrt 2$ are $3\times 3$
matrices, $s_i$ and $p_i$ stand for the \S\  and \P\  nonets and
$\lambda_i$ are the Gell-Mann matrices, normalized as $\Tr[\lambda_i\lambda_j]=
2\delta_{ij}$, and where for the singlet 
$\lambda_0 =\sqrt \frac 2 3 {\bf 1}$ is 
included. Note that each meson from the start has a definite $SU3$ symmetry
content, which in the quark model means that it has a definite $q\bar q$
content.

 Apart from the term ${\cal L}^{SB}$, 
Eq.(1) is clearly invariant under $\Sigma \to U_L\Sigma U_R^\dagger$.
In Eq. (1)  I have contrary to the usual convention defined the sign of $m^2$
such that the naive physical squared mass would be $-m^2$, and the instability
thus occurs when $m^2>0$.
Generally one includes in addition into $\cal{L}$ terms which break the 
symmeties:
\be
{\cal L}^{SB}=c \cdot [ {\rm det}\Sigma +
{\rm det}\Sigma^\dagger]+\epsilon_0 s_0 + \epsilon_8 s_8 \ .
\ee
 The $c$ term breaks the $U(1)$ symmetry
explicitely and gives the $\eta_1$ a mass, while 
$\epsilon_0$ gives the pseudoscalar octet mass and $\epsilon_8$
breaks explicitely SU3$_f $ \  down to isospin. 

As well known with  $-m^2<0$ and $\lambda>0, \lambda'>0 $
 the potential in Eq.(1)
(being of the form of a "Mexican hat") gives rise to an instability with vacuum
condensate $<s_0>=v=\sqrt (2/3) f_\pi$. 
Let $\lambda' =0$ for a moment, then shifting
as usual the scalar field, $\Sigma \rightarrow \Sigma + v$, one finds
$v =m^2/(4\lambda)$. 
Furthermore, the squared mass ($-m^2$) is for the \S\ nonet
replaced  by $-m^2+12\lambda v^2=2m^2$, while 
the \P\ nonet  becomes massless,
$-m^2+4\lambda v =0$. If we had included the $\lambda'$ term 
instead of the $\lambda $ term, then only the scalar singlet would aquire
mass while the remaining 17 states would be massless.

In addition, very importantly, after  shifting the scalar singlet field,  
the $\lambda$  term (keeping $\lambda'=0$) generates trilinear $spp$ and $sss$
couplings of the form $g[\Tr [\lambda_i \lambda_j \lambda_k +h.c.]/2$ and  
with an overall normalization $g=4\lambda f_\pi$. These couplings   obey
the simply connected, Okubo-Zweig-Iizuka  (OZI) allowed
quark line rules with flavour symmetry
exact. One has SU3$_f$ predictions relating different couplings constants.
Denoting by $\sigma$ the \uu\ scalar, and by $\sigma_s$ the \ss\ scalar one has
e.g. $g^2 =g^2_{\sigma\pi^0\pi^0}= \frac 1 2 g^2_{\sigma\pi^+\pi^-} 
=g^2_{\sigma K\bar K}=
\frac 2 3 g^2_{K^*_0K\pi}= g^2_{a_0K\bar K} =\frac 1 2 g^2_{\sigma_sK\bar K}$,
 $g_{\sigma_s\pi\pi}=0$ etc. Here we summed over charge states except for
the $\sigma\pi\pi$ couplings since conventionally the $\sigma\pi\pi$ coupling
is $g_{\sigma\pi^0\pi^0}$. If one includes also the $\lambda'$ term
of Eq.(1) then only the couplings involving the $\sigma$ and $\sigma_s$ 
states would be altered, which would violate the OZI rule at the tree level.

Such flavour symmetric OZI 
couplings, generated by the $\lambda$ term of Eq.(1), together with 
 a near-degenerate bare scalar nonet mass
were, in fact, the starting point of my recent analysis of the scalar $q\bar q$
nonet \cite{NAT}. In particular it was crucial that after determining the
overall coupling $g$ from a fit to data on 
 $K^*_0\to K\pi$ and $a_0\to\pi\eta$ one predicted correctly the $\pi\pi$ 
phase shifts. This shows that the above relations relating  the
bare $\sigma$ and $\sigma_s$ couplings to the same overall $g$ as those of
$K^*_0K\pi$ and $a_0\pi\eta$ must be approximately
satisfied experimentally. I.e., one cannot tolerate very big $\lambda'$ coupling
in Eq.(1) since then these relations would be  destroyed. 
Another argument for
that $\lambda'$ must be small is that then the bare scalar singlet
 state would have a very different mass from the other nonet members, 
not needed in Ref.\cite{NAT}.

 After the unitarization was 
performed in \cite{NAT} the scalars aquire finite widths
and are strongly shifted in mass by the different couplings to the \P \P\
thresholds. The \P\  masses in the thresholds were given their experimental
values, and consequently the main source of flavour symmetry breaking in the
output physical mass spectrum was generated by the vastly different positions
of these thresholds. E.g. the large experimental splitting between the $a_0(980)$
and $K^*_0(1430)$ masses came from the large breaking in the sum of
loops for the $K^*_0$ ($K^*_0\to K\pi , K\eta , K\eta' \to K^*_0$) compared to
those for $a_0$ ($a_0\to \pi\eta , K\bar K , \pi \eta' \to a_0$),
although in the strict SU3$_f$
limit these loops gives the same shift to the two resonances.

There were only 6 parameters in \cite{NAT} out of which two parametrized the
bare scalar spectrum (1.42 GeV for the $u\bar u$, $d\bar d$, $u\bar d$,
$d\bar u$, and an extra 0.1 GeV when an $s$ quark replaces a $u$ or $d$
quark). 
The overall coupling was parametrized by $\gamma=1.14$ and $k_0=0.56$ GeV/c
was the cut off parameter. Now, the $\gamma$ parameter can be related to $\lambda$
of Eq.(1) through $\lambda=2\pi\gamma^2 =8.17$, by comparing
the $\sigma \pi\pi$ coupling of the two schemes. [One has
$g^2_{\sigma\pi^0\pi^0}/(4\pi)$ $ = \gamma^2 m^2_{\sigma}= \lambda
m^2_{\sigma}/(2\pi)$. The latter equation follows from $g_{\sigma\pi^0\pi^0}
=g=4v\lambda$  and $v^2=-m^2/(4\lambda )=m_\sigma^2/(8\lambda)$ given
above.]
                                                
 Using the conventional value for $f_\pi=\sqrt(2/3) v=93$ 
MeV and $\lambda=8.17$
one predicts from (1) at the tree-level
 that the average mass of the $u\bar u,\ d\bar d,\ d\bar u$ states
is $m=v (8\lambda )^{1/2} $=920 MeV. Now although
strictly speaking there is no exact one-one correspondence between the
 model of Eq. (1) at the tree level, before renormalization, and the
unitarised model of \cite{NAT}, it is remarkable that this prediction
is close to the average of the  $\sigma$ and $a_0(980)$ masses found in
\cite{NAT}. I.e. if we had used $f_\pi=93$ MeV to determine the energy  scale,
one could have eliminated one of the 6 parameters.
  
\begin{figure}[h]
\centerline{
\protect
\hbox{
\psfig{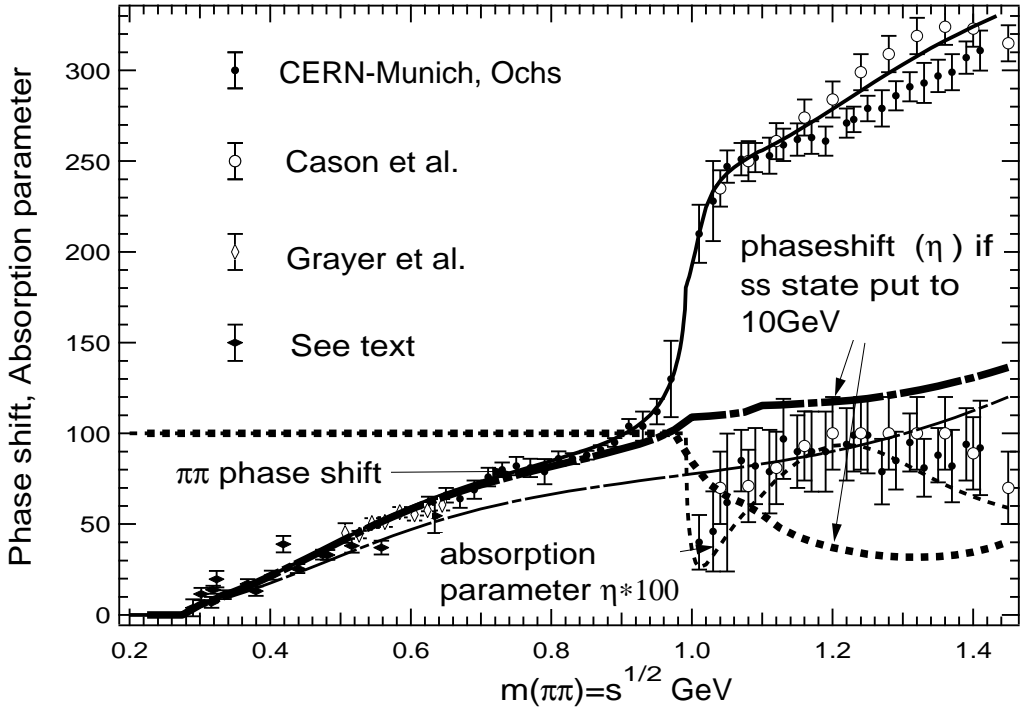}}}
\caption{The full line curve shows the 
I=0, phase shifts for $\pi\pi\rightarrow \pi\pi$,
as predicted by the  model [1].
 If one lets the $s\bar s$ meson mass
become very large (10 GeV)  in the model then the rapid rise at 1 GeV
due to the $f_0(980)$ vanishes. This shows that $f_0(980)$ is the unitarized
$s\bar s$ state. The remaining phase shift (dot-dashed thick line) is
then entirely due to the $u\bar u +d\bar d$ channel and one sees that what
remains can be understood as a very broad resonance. This is  the "$\sigma$", 
whose 90$^\circ$ mass value is at 880 MeV, while the pole is at 470-i250 MeV.
(The thin dot-dashed curve shows the phase shift if also the $K\bar K$ and
$\eta\eta$ thresholds are given large values.) }
\end{figure}                                    

Thus an imortant  point which I did not realize when writing \cite{NAT}
 is that that model  can be
interpreted as an effective field theory  given by ${\cal L}$ of Eq.(1)
with $\lambda=8.14$ and $\lambda'\approx 0$.  The absence of the $\lambda'$
term at the tree level means that the OZI rule holds exactly before
unitarization.
 The three small additional terms in  ${\cal L}^{SB}$  in Eq.(2) 
give the pseudoscalars their 
physical masses, and account for the extra 0.1 GeV for the bare strange 
states.

Of course the unitarization procedure of \cite{NAT} should be improved upon,
including $u$- and $t$-channel singularities \cite{isgur}, more loops etc., 
but I believe that the dominant effects were
already included phenomenologically for the scalar states. 
                                                
An important point observed in the second paper of Ref. \cite{NAT} was that
the model requires the existence of the light and broad $\sigma$ resonance.
This is explained in more detail in Fig.1 above.
Both before and at this conference there has appeared 
several new  papers \cite{ishi}, 
which through different analyses and models  support this same  conclusion, i.e.
that the light and broad sigma, which has been controversial for so long,
really exists, and is here to stay \cite{PDG}. 

Finally I have too little time left to explain sufficiently another recent
argument   
of mine \cite{NATPL},
which certainly many people would consider  speculative. It involves the
possibility that that the spontaneous breaking of symmetries does not 
stop after the standard step discussed above. After quantum loops further
symmetry breaking can occur, but here I refer interested to \cite{NATPL}
and papers in progress.

\end{document}